\documentclass[prb,twocolumn,amsmath,amssymb,letterpaper,superscriptaddress,nofootinbib]{revtex4-2} 
\usepackage{graphicx}
\usepackage{dcolumn}
\usepackage{float}
\usepackage{booktabs}
\usepackage[hypertexnames, breaklinks=true, bookmarksnumbered=true, bookmarksopen=true, colorlinks=true, linktocpage=true, citecolor=blue, urlcolor=magenta, linkcolor=magenta]{hyperref}
\usepackage{amsmath,amssymb,amsfonts,amsthm}
\usepackage{mathtools}
\usepackage[T1]{fontenc}
\usepackage[latin9]{inputenc}
\usepackage{txfonts}
\usepackage{bbm}
\usepackage{bm}
\usepackage{subfigure}
\usepackage{mathrsfs}
\usepackage{xcolor}
\usepackage{natbib}
\usepackage{siunitx}

\usepackage{appendix}
\begin{document}

\title{Dynamic transition of the density-matrix topology under parity-time symmetry}

\author{Wenzhi Wang}
 \affiliation{CAS Key Laboratory of Quantum Information, University of Science and Technology of China, Hefei 230026, China}
\author{Wei Yi}
\email{wyiz@ustc.edu.cn}
\affiliation{CAS Key Laboratory of Quantum Information, University of Science and Technology of China, Hefei 230026, China}
\affiliation{Anhui Province Key Laboratory of Quantum Network, University of Science and Technology of China, Hefei 230026, China}
\affiliation{CAS Center For Excellence in Quantum Information and Quantum Physics, Hefei 230026, China}
\affiliation{Hefei National Laboratory, University of Science and Technology of China, Hefei 230088, China}

\begin{abstract}
Density-matrix topology, defined through the geometric property of the relevant modular Hamiltonian, can undergo transitions in the corresponding open-system dynamics.
While symmetry considerations are crucial to ensure such a dynamic topological transition, we show that a hidden parity-time symmetry can further facilitate it.
Considering the Lindbladian dynamics of a fermionic Gaussian state, we extract the time-evolved density-matrix topology from the single-particle correlation, whose dynamics is governed by a non-Hermitian damping matrix.
We show that, for a parity-time symmetric damping matrix and a chiral symmetric correlation matrix, a dynamic transition in the density-matrix topology necessarily occurs in the parity-time unbroken regime where eigenvalues of the damping matrix are real.
We illustrate our results using a concrete model, and map out the dynamic phase diagram.
Remarkably, we find that the dynamic transition can also happen periodically in the parity-time symmetry broken regime.
\end{abstract}

\maketitle

\section{Introduction}\label{sec1}

Topology is an extensively researched paradigm in modern quantum physics.
Whether it be the global topological features of the ground-state wave function in the Hilbert space~\cite{HKrmp10,QZrmp11}, or the emergent dynamic structures in the synthetic (often time-related) manifolds~\cite{Levin13,Bhaseen15,Refael16,Zhai17,cooper18,Xiong-Jun,Chen17,Ueda17,wytheory18,Weitenberg17,shuaichen,pengxue19}, topological matters and topological phenomena are universally characterized by integer-valued invariants. They are robust against local perturbations, and promise useful applications for quantum transport and control.
While topology is routinely defined and studied for pure states, in quantum many-body systems, one is often confronted with situations where a mixed-state description is inevitable. Indeed, by introducing geometric phases and topological invariants for mixed states, the study of density-matrix topology has stimulated significant interest in recent years~\cite{delgado13,delgado14,Uhl,EGP,GJ,fPRB,fPRL}.
An outstanding example is the ensemble geometric phase, which, in one-dimensional systems, represents a natural generalization of the Zak phase, and is in principle detectable through many-body interferometry or quantized transport~\cite{EGP,fPRB}. The corresponding topological invariant is
determined by the topology of the modular Hamiltonian $\hat{K}$, defined through the density matrix as
$\hat{\rho}=e^{-\hat{K}}$. Under such a definition, the density-matrix topology is protected by the spectral gap of $\hat{K}$, and only changes when the gap closes.
While topological transitions of this kind can emerge for systems at thermal equilibrium, they also arise in the dynamics of open systems. For instance, a recent study shows that, based on symmetry considerations, dynamic transitions of the density-matrix topology can be engineered in the Lindbladian dynamics~\cite{zhai23}.

For quantum open systems governed by the Lindblad master equation~\cite{opensys,Nonh}, their intrinsic non-Hermiticity further lends themselves to fresh insights from the perspectives of non-Hermitian physics.
On one hand, by vectorizing the density matrix, the Liouvillian is mapped to a non-Hermitian matrix in an enlarged Hilbert space~\cite{tyson03,vidal04,yineng}. On the other hand, information of the density-matrix dynamics is reflected in the evolution of the single-particle correlation, driven by a non-Hermitian damping matrix~\cite{diehlnp,Cevo}.
Within either framework, unique features of non-Hermitian matrices, such as the parity-time (PT) symmetry and exceptional points~\cite{pt1,pt2,pt5,pt6,pt7,ep3,ep4,eptopo}, the non-Hermitian topology~\cite{gongprx,nonHtopo1,nonHtopo2} and the non-Hermitian skin effect~\cite{WZ1,Budich,BBK21,LTL23}, can manifest themselves in quantum open systems through the damping matrix, impacting both the transient and the long-time dynamics~\cite{Cevo,LEP1,LEP2,LEP3,sq1,sq2,qhe,sunyi,rydbergexp,cavityskin}.
In this work,  we show that this is also the case with the density-matrix topology.

Focusing on a fermionic Gaussian state, which corresponds to a mixed state of non-interacting fermions~\cite{rmpgauss}, we find that the dynamic transition of the density-matrix topology can be facilitated by the PT symmetry of the damping matrix.
For a Gaussian state, the single-particle correlation provides a complete description of the density matrix, and is therefore connected to the modular Hamiltonian through a closed-form expression.
It follows that the non-Hermitian damping matrix, which governs the time evolution of the single-particle correlation, provides a natural connection between non-Hermitian features and density-matrix topology.
Specifically, through general analysis and a generic example, we demonstrate that, for a PT symmetric damping matrix and a chiral symmetric correlation matrix, a dynamic transition in the density-matrix topology necessarily occurs in the PT-symmetry unbroken regime, where eigenvalues of the damping matrix are real.
Beyond such a regime, the dynamic transition occurs in a model- and parameter-dependent manner. Intriguingly, in our example, the dynamic transition can happen periodically in the PT-symmetry broken regime, when the density-matrix approaches the vacuum state at long times.
Our work reveals the close connection between the mixed-state topology in quantum open systems and the unique features of non-Hermitian physics.

The work is organized as follows. In Sec.~II, we review basic conceptions of Gaussian states of fermions, quadratic Lindblad equations and the chiral symmetry, and we define the topological invariant. We then consider the PT symmetry of the damping matrix and discuss its effect on the evolution of topological properties of the system in Sec.~III. In Sec.~IV, we discuss a concrete example. Finally, We summarize in Sec.~V.

\section{General framework}\label{sec2}

We consider the evolution of a Gaussian state of fermions under a quadratic Lindblad equation
\begin{equation}
\frac{d\hat{\rho}}{dt}=-i[\hat{H},\hat{\rho}]+\sum_{\mu}(2\hat{L}_{\mu}\hat{\rho}\hat{L}_{\mu}^{\dagger}-\{\hat{L}_{\mu}^{\dagger}\hat{L}_{\mu},\hat{\rho}\}),\label{eq:lindblad}
\end{equation}
where $\hat{H}=\sum_{ij}\hat{c}_{i}^{\dagger}H_{ij}\hat{c}_{j}$ is quadratic in terms of the fermionic creation (annihilation) operators $c^\dag_i$ ($c_i$). Here, $H_{ij}$ are the elements of matrix $H$, which denotes the Hamiltonian in the single-particle basis.
For concreteness, we consider a one-dimensional lattice model, so that the subscript $i$ consists of both the unit-cell and the internal-state (or sublattice) degrees of freedom, denoted as $(x,s)$. The linear quantum jump operators are given by $\hat{L}_{\mu}=\sum_{i}D_{\mu i}\hat{c}_{i}$, where $D_{\mu i}$ are the complex coefficients. Note that we only consider lossy jump operators given by superpositions of annihilation operators, which are relevant to dissipative quantum open systems.

For a Gaussian mixed state, its modular Hamiltonian is given by $\hat{K}=\sum_{ij}\hat{c}_{i}^{\dagger}K_{ij}\hat{c}_{j}$, where $K_{ij}$ is identified as the element of the modular Hamiltonian matrix $K$.
Our study is based on two important features of the Gaussian state.
First, under a quadratic Lindblad equation, the time evolution of an initial Gaussian state keeps its Gaussian form. Second, the single-particle correlation matrix, with elements $C_{ij}=\text{Tr}(c^\dag_i c_j\hat{\rho})$, fully captures the information of the Gaussian mixed state. It follows that, at any given time during the density-matrix dynamics, the correlation matrix $C$ and the modular Hamiltonian matrix $K$ are related through (see Appendix \ref{appA})
\begin{equation}
C=\frac{1}{e^{K^{T}}+1}.\label{eq:CK}
\end{equation}
From the Lindblad equation Eq.~(\ref{eq:lindblad}), we can find the evolution equation of $C$ (see Appendix \ref{appB})
\begin{equation}
\frac{dC}{dt}=XC+CX^{\dagger},\label{eq:CX}
\end{equation}
where the damping matrix $X=iH^T-M^T$, with the matrix $M$ defined through its elements
$M_{ij}=\sum_{\mu}D_{\mu i}^{*}D_{\mu j}$.
Apparently, the dynamics of the correlation matrix $C$ is governed by the non-Hermitian damping matrix $X$. Exotic features of the non-Hermitian physics may then manifest in open-system dynamics through the damping matrix. Since our goal is to study the topological properties of $K$, according to Eq.~(\ref{eq:CK}), we focus on $C^T$ in the following discussions for convenience. we further define $\tilde{X}=X^\ast$, so that the evolution of $C^T$ is driven by $\tilde{X}$.

Finally, under the lattice translational symmetry, we derive the equation of motion for the correlation matrix in the quasi-momentum space. This is achieved using the Bloch states $|k\rangle=\sum_{x}e^{ikx}|x\rangle$, and by writing a general matrix $O$ as
\begin{equation}
O=\sum_{k} O_k \otimes |k\rangle \langle k|,\label{eq:CkXk}
\end{equation}
where $O_k$ is a matrix in the basis of internal states (or sublattices), and $O$ could be any one of
the matrices $K$, $C^T$, $H$, $M$, or $\tilde{X}$.
We then have
\begin{equation}
\frac{d C^T_{k}}{dt}=(\tilde{X}_{k}C^T_{k}+C^T_{k}\tilde{X}_{k}^{\dagger}),\label{eq:kspaceCX}
\end{equation}
and Eq.~(\ref{eq:CK}) can be cast into the quasi-momentum space
\begin{equation}
C^T_k=\frac{1}{e^{K_k}+1}.\label{eq:kspaceCK}
\end{equation}

According to Eq.~(\ref{eq:kspaceCK}), the density-matrix topology, if any, is defined through $K$ and encoded in $C^T$. In the following, we focus on matrices $K$ with chiral symmetry, so that a winding number $\nu$ can be defined as the topological invariant.
As an illustrating example, we consider a two-dimensional internal-state subspace (or sublattice space), and write
\begin{align}
K_k=\alpha_{K}(k) \sigma_0+\bm{n}_{K}(k)\cdot \bm{\sigma},\label{eq:Kk}
\end{align}
where $\alpha_{K}(k)$ is a $k$-dependent real number, $\bm{n}_{K}(k)$ is a $k$-dependent real vector, $\bm{\sigma}$ is a three-dimensional vector consisting of Pauli matrices, and $\sigma_0$ is a $2\times 2$ identity matrix.
The topology
of $K$ is determined by its eigenvectors, and independent of $\alpha_K(k)$.
We therefore consider the chiral symmetry
\begin{equation}
\Gamma (\bm{n}_{K}\cdot \bm{\sigma})\Gamma^{-1}=-\bm{n}_{K}\cdot \bm{\sigma},\label{eq:chiral}
\end{equation}
where the symmetry operator $\Gamma=\bm{n}_\Gamma\cdot \bm{\sigma}$, with the unit vector $\bm{n}_\Gamma$ denoting the chiral symmetry axis. The winding number is then
\begin{equation}
\nu=\frac{1}{2\pi}\oint (\bm{n}_{K}^{u}\times\frac{d\bm{n}_{K}^{u}}{dk})_\Gamma dk.\label{eq:winding}
\end{equation}
where $\bm{n}_{K}^{u}=\bm{n}_{K}/|\bm{n}_{K}|$, and ${(\cdots)}_{\Gamma}$ indicates a projection
on to the chiral axis. Geometrically, the chiral symmetry requires that $\bm{n}_{K}$ lies in the plane perpendicular to $\bm{n}_\Gamma$ (see Appendix \ref{appC}).
We note that the chiral symmetry above is more general than the one considered in Ref.~\cite{zhai23}, where a more stringent requirement $\Gamma K_k\Gamma^{-1}=-K_k$ is adopted. While the latter necessitates $\alpha_{K}=0$, the two cases correspond to different physical states, since $K_k$ with $\alpha_{K}=0$ and $\alpha_{K}\neq 0$ yield different correlation matrices. Furthermore, as we illustrate later, when only lossy jump operators are considered, $\alpha_K\neq 0$ generally holds, necessitating the definition of the generalized chiral symmetry to characterize the density-matrix topology.

\begin{figure}[tbp]
\includegraphics[width=8cm]{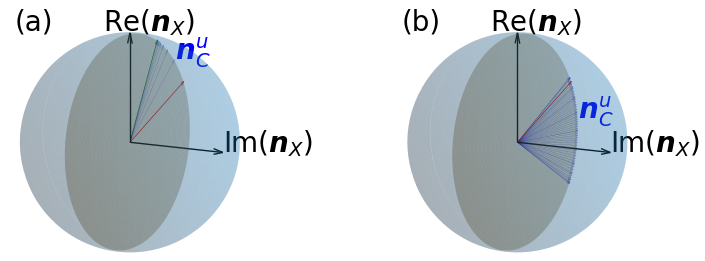}
\caption{Schematic illustrations of the correlation matrix for a given $k$, (a) in the PT unbroken regime and (b) in the PT broken regime.
In either figure, $\text{Im}(\bm{n}_{X})$ indicates the chiral axis, the shaded plane is perpendicular to the chiral axis. The correlation matrix is represented by the normalized vector
$\bm{n}_{C}^{u}$, shown as blue arrows.
The red arrows in (a) and (b) indicate the initial $\bm{n}_{C}^{u}$,
and the green arrow in (a) indicates the steady-state $\bm{n}_{C}^{u}$ in the long-time limit.
\label{fig1}}
\end{figure}

\section{PT-symmetric damping matrix.}\label{sec3}
The non-Hermiticity of the damping matrix opens up the avenue of investigating the impact of non-Hermitian physics on the density-matrix topology. As an illustration, in this study, we consider a damping matrix with PT symmetry. Being a two-dimensional matrix, a PT symmetric $\tilde{X}_k$ is equivalent to its pseudo-Hermiticity~\cite{Nonh}. Specifically, there exists a two-dimensional invertible Hermitian matrix $\eta_k$, such that
$\tilde{X}_{k}^{\dagger}=\eta_{k}\tilde{X}_{k}\eta_{k}^{-1}$.
Assuming the existence of a complete biorthogonal eigenbasis,
the pseudo-Hermiticity of matrix $\tilde{X}_{k}$ is equivalent to the condition that
its eigenvalues are either real or in complex conjugate pairs~\cite{Nonh}.
Generally, a pseudo-Hermitian $\tilde{X}_k$ can be written as
\begin{equation}
\tilde{X}_{k}=\alpha_{X}(k)\sigma_{0}+\bm{n}_{X}(k)\cdot\bm{\sigma},\label{eq:ptXk}
\end{equation}
where $\alpha_{X}(k)$ is a $k$-dependent real number, and $\bm{n}_{X}(k)$ is
\begin{equation}
\bm{n}_{X}(k)=\gamma_{k}\bm{n}_{1k}+i\rho_{k}\sin\theta_{k}\bm{n}_{2k}+i\rho_{k}\cos\theta_{k}\bm{n}_{3k},\label{eq:ptnXk}
\end{equation}
where $\gamma_k,\rho_k,\theta_k\in \mathbb{R}$, and $\bm{n}_{jk}$ ($j=1,2,3$) are three orthogonal vectors. Here $\tilde{X}_k$, $\alpha_{X}$, and $\bm{n}_{X}$ are all time-independent.


We are interested in the case that the density-matrix topology is well-defined throughout the time evolution (except for the transition points). This requires that the chiral symmetry of $C^T$ is not broken throughout the dynamics, which imposes additional constraints on $\tilde{X}_k$.
We denote the form of $C^T$ as
\begin{equation}
C^T_k=\alpha_{C}(k)\sigma_0+\bm{n}_{C}(k)\cdot \bm{\sigma},\label{eq:Ck}
\end{equation}
where $\alpha_{C}(k)$ is a $k$-dependent positive number, and $\bm{n}_{C}(k)$ is a $k$-dependent real vector. Here $C^T_k$, $\alpha_C$, and $\bm{n}_C$ are time-dependent.
Correspondingly, $K_k$, $\alpha_{K}$, and $\bm{n}_{K}$ are also time-dependent.
Since  $\bm{n}_{C}$ and $\bm{n}_{K}$ are parallel but opposite in direction for all $k$ (see Appendix \ref{appC}), the winding number is unchanged when replacing $\bm{n}_{K}^{u}$ in Eq.~(\ref{eq:winding}) with $\bm{n}_{C}^{u}=\bm{n}_{C}/|\bm{n}_{C}|$.
Considering the equation of motion Eq.~(\ref{eq:kspaceCX}), it follows that the chiral symmetry is kept unbroken throughout the time evolution, provided
$\text{Im}(\bm{n}_{X})$ is parallel to $\bm{n}_{\Gamma}$ for all $k$ (see Appendix \ref{appC}). Note that this is only a sufficient condition.

Under such a premise and without loss of generality, we take $\bm{n}_{3k}=0$ and $\Gamma=\sigma_z$ in the following. We then have
$\tilde{X}_k:=-iH_k-M_k=-i\rho_k\sigma_z+(\alpha_{X}\sigma_{0}+\gamma_{k}cos\theta_{k}\sigma_{x}+\gamma_{k}sin\theta_{k}\sigma_{y})$.
The eigenvalues of $\tilde{X}_{k}$ are then given by
\begin{equation}
\varepsilon_{\pm k}=\alpha_{X}\pm \sqrt{\gamma_{k}^{2}-\rho_{k}^{2}}.
\end{equation}
Denoting the left and right eigenstates of $\tilde{X}_k$ as $|L_{\pm k}\rangle$ and $|R_{\pm k}\rangle$, and using the biorthonormal condition
$\langle L_{\xi k}|R_{\xi^{\prime} k}\rangle=\delta_{\xi \xi^{\prime}}$ (with $\xi,\xi^{\prime}=\pm$), we derive the time-evolved correlation matrix from Eq.~(\ref{eq:kspaceCX}) (see Appendix \ref{appD})
\begin{align}
C^T_{k}(t)=
\sum_{\xi,\xi^{\prime}=\pm}e^{(\varepsilon_{\xi k}+\varepsilon_{\xi^{\prime}k}^{*})t}|R_{\xi k}\rangle\langle R_{\xi^{\prime}k}|\langle L_{\xi k}|C^T_{k}(0)|L_{\xi^{\prime}k}\rangle.\label{eq:Ct}
\end{align}

We are now in a position to discuss the impact of PT symmetry on the correlation-matrix dynamics.
First, given its definition, $M=D^\dag D$ is semi-positive definite. Here the elements of the coefficient matrix $D$ are $D_{\mu i}$. Since both $H_k$ and $M_k$ are Hermitian, we have
$M_k=-\alpha_{X}\sigma_{0}-\gamma_{k}cos\theta_{k}\sigma_{x}-\gamma_{k}sin\theta_{k}\sigma_{y}$,
and it follows that $\alpha_{X} \leq 0$ and $|\alpha_{X}|\geq |\gamma_{k}|$, so that $\text{Re}(\varepsilon_{\pm k})\leq0$. For any given $k$, if $\gamma_{k}^{2}>\rho_{k}^{2}$, then $\varepsilon_{\pm k}$ are real and $\varepsilon_{\pm k}\leq0$, meaning the PT symmetry is unbroken in the corresponding $k$-sector.
It follows that, in the long-time limit, the dominant contribution on the right-hand side of Eq.~(\ref{eq:Ct}) comes from the exponential factor $e^{(\varepsilon_{+ k}+\varepsilon_{+k}^{*})t}$.
Hence, the correlation matrix evolves toward $|R_{+k}\rangle\langle R_{+k}|$ with a vanishing amplitude in the long-time limit.
This is schematically illustrated in Fig.~\ref{fig1}(a), where the correlation-matrix evolution is represented by that of the normalized vector $\bm{n}^u_C$ on the Bloch sphere.
By contrast, when $\gamma_{k}^{2}<\rho_{k}^{2}$, $\varepsilon_{\pm k}$ are a complex conjugate pair, and the PT symmetry is broken. Within such a $k$-sector, the correlation matrix is oscillatory in time, but also with a vanishing amplitude in the long-time limit.
As illustrated in Fig.~\ref{fig1}(b), this is represented by a periodic evolution of $\bm{n}^u_C$ within the plane perpendicular to the chiral axis.
We further note that, according to Eq.~(\ref{eq:kspaceCX}), the time evolution of $C^{T}_{k}$
is generally not trace-preserving.
From Eqs.~(\ref{eq:CK}) and (\ref{eq:Kk}), we then have $\alpha_K\neq 0$ in genereal.
This is because, if $\alpha_K$ remains $0$ during the evolution, we have $\text{Tr} (K_k)=0$, leading to $\text{Tr} (C^T_k)=1$.
Note that this is  in sharp contrast with Ref.~\cite{zhai23}, where $\alpha_K=0$ is achieved at the cost of additional constraints on the gain and loss jump operators.



\begin{figure*}[tbp]
\centering
\includegraphics[width=\linewidth]{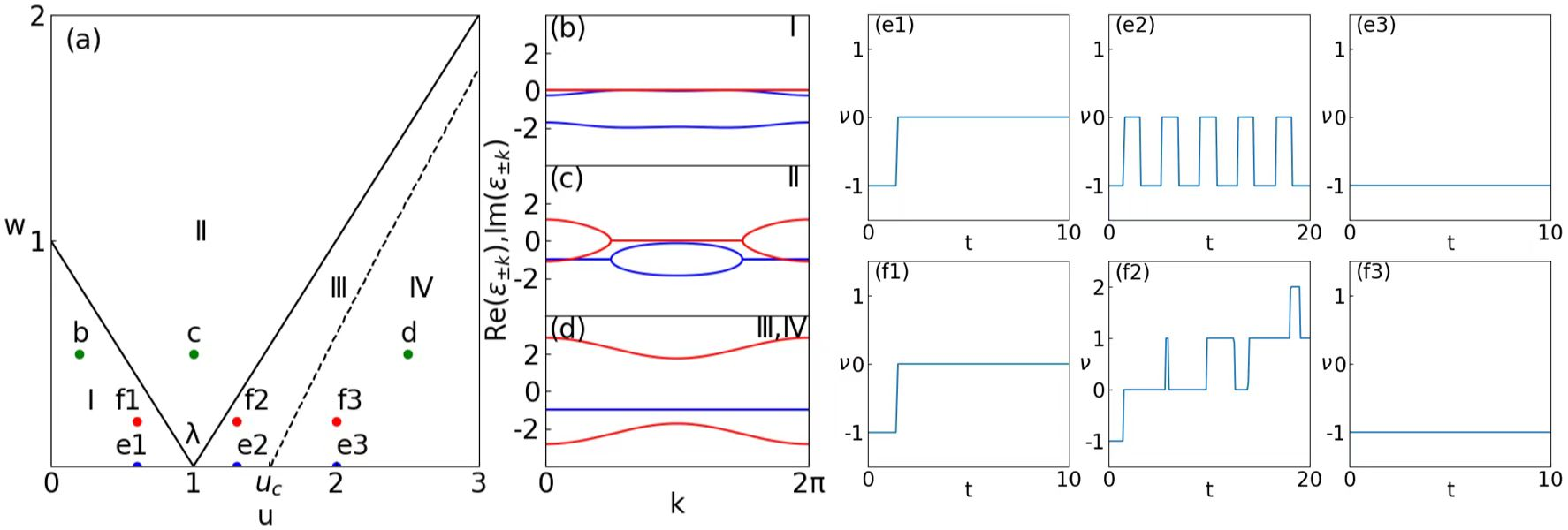}
\caption{
(a) Phase diagram for the PT symmetry and dynamic transitions of model Eq.~(\ref{eq:model}).
Region I: PT-symmetry fully-unbroken region, with typical eigenvalues of the damping matrix $\epsilon_{\pm k}$ illustrated in (b), where $u=0.2,w=0.5$. In this region, the dynamic transition occurs at least once [see (e1)(f1)].
Region II: PT-symmetry partially broken region, with typical $\epsilon_{\pm k}$ illustrated in (c), where $u=1,w=0.5$.
In this region, the behavior of the dynamic transition is a crossover between those of regions I and III.
Region III and IV: PT-symmetry fully broken region, with typical $\epsilon_{\pm k}$ illustrated in (d), where $u=2.5,w=0.5$. In region III, the dynamic transition can occur repeatedly [see Fig.~\ref{fig2}(e2)(f2)]. In region IV, the dynamic transition does not occur [see Fig.~\ref{fig2}(e3)(f3)]. In (b)(c)(d), blue (red) lines represent the real (imaginary) parts of $\epsilon_{\pm k}$. (e1)(e2)(e3) Dynamic transitions in the flat-band limit, with $w=0$ and (e1) $u=0.6$, (e2) $u=1.3$, (e3) $u=2$.
(f1)(f2)(f3) Dynamic transitions beyond the flat-band limit, with $w=0.2$ and (f1) $u=0.6$, (f2) $u=1.3$, (f3) $u=2$.
For all calculations, we fix $\lambda=1$, $a=1$, and $b=2$.
\label{fig2}}
\end{figure*}

From Eq.~(\ref{eq:winding}), the density-matrix topology is reflected in the winding of $\bm{n}_{C}$ as $k$ traverses the first Brillouin zone with $k\in (-\pi, \pi]$. Hence, the sufficient conditions for a topological transition of the density matrix to occur in the open-system dynamics is as follows:
(i) the PT symmetry is unbroken for all $k$, a condition referred to as the PT fully unbroken regime in the literature;
(ii) the initial density matrix is topologically distinct (featuring different winding numbers) from the that in the long-time limit.
Here condition (i) ensures that the system in the long-time limit relaxes to a steady state, with the correlation matrix in each $k$-sector given by $|R_{+k}\rangle\langle R_{+k}|$.
Since the density-matrix topology is well-defined throughout the dynamics, condition (ii) then necessitates at least one transition in the density-matrix topology as the steady-state is approached. Hence, PT symmetry of the damping matrix facilitates the dynamic transition of the density-matrix topology, by ensuring the existence of a definite topology of the final steady state.
While the situation is less clear beyond the PT-symmetry unbroken regime, we show in the following that an interesting, model-dependent scenario can arise, where the dynamic transition of the density-matrix topology occurs periodically.

\section{An example}\label{sec4}
We illustrate the discussions above using a generalized Su-Schieffer-Heeger model, where the coherent Hamiltonian is given by
\begin{align}
\hat{H}=
u\sum_{x}\hat{c}_{x,A}^{\dagger}\hat{c}_{x,B}+\frac{1}{2}w\sum_{x}(\hat{c}_{x+1,A}^{\dagger}\hat{c}_{x,B}+\hat{c}_{x,A}^{\dagger}\hat{c}_{x+1,B})+H.c..
\label{eq:model}
\end{align}
Here $A$ and $B$ represent the sublattice sites, $x$ is the unit-cell label, and $u,w\in \mathbb{R}$ are hopping rates. The quantum jump operators are given by $\hat{L}_{x}=\sqrt{2\lambda}\hat{c}_{x,B}$ ($\lambda\in \mathbb{R}$), depicting local single-particle loss.
This leads to a PT symmetric damping matrix (in the $k$-space)
\begin{align}
\tilde{X}_{k}=\begin{bmatrix}0 & -i(u+w\cos k)\\
-i(u+w\cos k) & -2\lambda
\end{bmatrix},\label{eq:Xkex}
\end{align}
with eigenvalues $\varepsilon_{\pm k}=-\lambda\pm \sqrt{\lambda^{2}-(u+wcosk)^{2}}$. Here a PT symmetry fully unbroken regime exists for $u+w< \lambda$, where the eigenvalues $\varepsilon_{\pm k}$ are real for all $k$. This is indicated in Fig.~\ref{fig2}(a) as region I.
When $u-w>\lambda$, the PT symmetry is fully broken where the eigenvalues are complex for all $k$.
This is indicated in Fig.~\ref{fig2}(a) as regions III and IV.
In between these regions, the PT symmetry is unbroken in some $k$-sectors, and broken in others, as exceptional points emerge in the $k$ space.
This corresponds to region II in Fig.~\ref{fig2}(a). Typical eigenvalues of the damping matrix in the $k$ space are shown in Fig.~\ref{fig2}(b)(c)(d) for the different regions discussed above.
Importantly, in the PT-symmetry fully unbroken regime, the steady-state correlation matrix features a trivial topology, with a vanishing winding number. Whereas in other regimes, the correlation function does not have a global steady state, because of the presence of PT-symmetry broken $k$-sectors.

Next we examine the evolution of $C^T$, starting from an initial Gaussian state
\begin{align}
C^T_k(0)=\frac{1}{2}\left[1-\frac{\tanh(|\bm{n}_k|)}{|\bm{n}_k|}\bm{n}_k\cdot\bm{\sigma}\right],\label{eq:Ckini}
\end{align}
where $\bm{n}_k=(0,b\sin k,a+b\cos k),a>0,b>0$. The winding number of the initial state is calculated by normalizing $\bm{n}_k$ and replacing it for ${\bm{n}_K^u}$ in Eq.~(\ref{eq:winding}). Geometrically, the winding number counts the number of loops the vector $\bm{n}_k$ cycles around the origin as $k$ traverses the range $[0,2\pi)$.
Apparently, we have $\nu=1$
$(\nu=0)$ if $a<b$ $(a>b)$.
Following the previous discussions, we choose an initial state with $a<b$, so that it is topologically non-trivial. In the PT-symmetry fully unbroken regime, a topological transition should occur before the system reaches the topologically trivial steady state.

For a more systematic understanding of the dynamic transition, we start by fixing $\lambda=1$, and analyzing the flat-band limit with $w=0$ where the eigenvalues of the damping matrix are independent of $k$.
Here an exceptional point exists at $u=1$ for all $k$. To the left of the exceptional point ($u<1$), the PT symmetry is unbroken for all $k$, and a topological transition of the density matrix occurs in the steady-state approaching dynamics. This is illustrated in Fig.~\ref{fig2}(e1).
Past the exceptional point to the right, a global steady state does not exist, and it is difficult to establish a general condition for the occurrence of topological transition. Nevertheless, we numerically find that, for $u>u_c=1.53$, no dynamic transitions can be found, as illustrated in Fig.~\ref{fig2}(e3). We note that the location of $u_c$ is model dependent.
For $1<u<u_c$, on the other hand, dynamic transitions can either occur periodically [see Fig.~\ref{fig2}(e2)], or not at all, depending on the initial condition characterized by  $a$ and $b$.
Here the PT symmetry is fully broken, meaning the eigenvalues of the damping matrix take the form of complex-conjugate pairs, and the correlation matrix evolves in an oscillatory fashion. In the flat-band limit, the oscillation is further synchronized for all $k$ sectors.
If it happens that a dynamic topological transition occurs in one cycle of the oscillation, then it should occur periodically. The periodic recurrence of the dynamic topological transition is therefore enforced by the fully broken PT symmetry and the flat-band condition.

Moving away from the flat-band limit, the three sectors discussed above evolve into regions I, III and IV in the phase diagram, respectively. Typical dynamic transitions in these regions are shown in
in Fig.~\ref{fig2}(f1)(f2)(f3). While the dynamic features in regions I and IV are similar to those in the flat-band limit, in region III, the dynamic transitions can still arise repeatedly, but no longer in a strict periodical manner [see Fig.~\ref{fig2}(f2)].
This is because the oscillations of the correlation matrices in different $k$-sectors are no longer synchronized.
Additionally, a PT-symmetry partially broken region (region II) emerges for $w\neq 0$, where eigenvalues are real only for some $k$ sectors as discrete exceptional points appear in the $k$ space [see Fig.~\ref{fig2}(c)]. In terms of the dynamic transition, region II can be seen as a crossover between regions I and III, manifesting features of either one depending on the parameters. Note that in region II, both the real and imaginary gaps close at the exceptional points. But since they are removable for the integral in Eq.~(\ref{eq:winding}), a global winding number can still be defined therein~\cite{nsr23}.

\section{Discussion}\label{sec5}
We have shown that dynamic transitions of the density-matrix topology can be facilitated by a hidden PT symmetry of the damping matrix.
The dynamic transition of the density-matrix topology necessarily occurs or exhibits oscillatory behaviors, depending on the spontaneous PT symmetry breaking.
We consider dissipative open systems with purely lossy quantum-jump processes in this study, which necessitates a more general chiral symmetry adopted in Ref.~\cite{zhai23}, and enables the design of PT symmetric damping matrices that preserve the chiral symmetry of the correlation-matrix evolution.
While it is generally difficult to experimentally measure the density-matrix topology or its dynamic transitions, it is expected that the transition can in principle be probed through the many-body interferometry~\cite{EGP}.
Our study represents a first attempt at demonstrating the manifestations of
outstanding features in the non-Hermitian physics in the density-matrix topology in many-body quantum open systems, thus bridging two active research directions.

Going forward, it would be interesting to explore the impact of general
topology of PT-symmetric matrices on density-matrix topology~\cite{yang24}, or the impact of PT symmetry on the density-matrix topology for multiband modular Hamiltonians in higher dimensions.
It would be interesting to investigate the implication of different topological classifications of non-Hermitian models on the density-matrix topology of open systems~\cite{gongprx,nonHtopo1,nonHtopo2}.
We also expect other non-Hermitian features, particularly topology related features such as the non-Hermitian skin effect~\cite{WZ1}, to have an impact on the density-matrix topology.
Another promising direction is the density-matrix topology and dynamic transitions for non-Gaussian states. Therein, higher-order correlation functions are needed to depict the open-system dynamics. The interplay of modular-Hamiltonian topology and non-Hermitian dynamics of these correlation functions is yet to be explored.

\begin{acknowledgments}
We thank Fan Yang for helpful comments and discussions. This work is supported by the National Natural Science Foundation of China (Grant No. 12374479) and the Innovation Program for Quantum Science and Technology (Grant No. 2021ZD0301200).
\end{acknowledgments}

\appendix

\renewcommand{\thesection}{\Alph{section}}
\renewcommand{\thefigure}{A\arabic{figure}}
\renewcommand{\thetable}{A\Roman{table}}
\setcounter{figure}{0}
\renewcommand{\theequation}{A\arabic{equation}}
\setcounter{equation}{0}

\section{Derivation of the relationship between $C$ and $K$}\label{appA}

In this section, we give a detailed derivation of Eq.~(\ref{eq:CK}). For convenience, we write the fermion operators in the vector form, $\psi={(\hat{c}_{1},\hat{c}_{2},...,\hat{c}_{N})}^{T}$, where $N$ is the total number of degrees of freedom of the system. Following this format, we write $C$ and $\hat{K}$ as
\begin{equation}
C=\text{Tr}[{(\psi^{\dagger})}^{T}{\psi}^{T}\hat{\rho}],\quad \hat{K}={\psi}^{\dagger}K{\psi}.\label{eq:vecCK}
\end{equation}
Notice that in Eq.~(\ref{eq:vecCK})
\begin{equation}
{\psi}^{\dagger}={(\hat{c}_{1}^{\dagger},\hat{c}_{2}^{\dagger},...,\hat{c}_{N}^{\dagger})},{\psi}^{T}={(\hat{c}_{1},\hat{c}_{2},...,\hat{c}_{N})},
\end{equation}
which means that under our formatting convention, the transpose conjugate symbol operates on both the fermion operators and the vector form, but the transpose symbol operates only on the vector form. In general, $K$ is not diagonal, and we introduce a $N\times N$ unitary matrix $U$ to diagonalize $K$
\begin{equation}
\hat{K}={\psi}^{\dagger}{U}^{\dagger}UK{U}^{\dagger}U\psi.
\end{equation}
We then define $\tilde{\psi}=U\psi$ and ${K}^{\prime}=UK{U}^{\dagger}$, where ${K}^{\prime}$ is diagonal. On the other hand, using $\tilde{\psi}$, we define
\begin{equation}
\begin{split}
C^{\prime}&=\text{Tr}[{({\tilde{\psi}}^{\dagger})}^{T}{\tilde{\psi}}^{T}\hat{\rho}]\\
&=\text{Tr}[{({\psi}^{\dagger}{U}^{\dagger})}^{T}{(\psi U)}^{T}\hat{\rho}]\\
&=\text{Tr}[{({U}^{\dagger})}^{T}{(\psi^{\dagger})}^{T}{\psi}^{T}{U}^{T}\hat{\rho}]\\
&={({U}^{\dagger})}^{T}\text{Tr}[{(\psi^{\dagger})}^{T}{\psi}^{T}\hat{\rho}]{U}^{T}\\
&={({U}^{\dagger})}^{T}C{U}^{T}.\label{eq:priC}
\end{split}
\end{equation}
Note that $C^{\prime}$ can be viewed as the correlation matrix corresponding to the diagonalized modular Hamiltonian matrix ${K}^{\prime}$. Since ${K}^{\prime}$ is a diagonal matrix, it is easy to find that the off-diagonal elements of $C^{\prime}$ vanish. On the other hand, the diagonal elements of $C^{\prime}$ can be obtained according to the Fermi-Dirac distribution. So we have
\begin{equation}
C^{\prime}=\frac{1}{1+e^{{K}^{\prime}}}=\frac{1}{(1+e^{{K}^{\prime T}})}.\label{eq:priCK}
\end{equation}
Substituting Eq.~(\ref{eq:priC}) into Eq.~(\ref{eq:priCK}), we have
\begin{equation}
\begin{split}
C&={U}^{T}\frac{1}{(1+e^{{K}^{\prime T}})}{({U}^{T})}^{\dagger}\\
&={[{({U}^{T})}^{\dagger}]}^{-1}{(1+e^{{K}^{\prime T}})}^{-1}{({U}^{T})}^{-1}\\
&={[{U}^{T}(1+e^{{K}^{\prime T}}){({U}^{T})}^{\dagger}]}^{-1}\\
&=\frac{1}{1+\text{exp}[{U}^{T}{K}^{\prime T}{({U}^{T})}^{\dagger}]} \\
&=\frac{1}{1+\text{exp}[{({U}^{\dagger}{K}^{\prime}U)}^{T}]}\\
&=\frac{1}{1+e^{{K}^{T}}},
\end{split}
\end{equation}
which reproduces Eq.~(\ref{eq:CK}).

\renewcommand{\thefigure}{B\arabic{figure}}
\renewcommand{\thetable}{B\Roman{table}}
\setcounter{figure}{0}
\renewcommand{\theequation}{B\arabic{equation}}
\setcounter{equation}{0}

\section{Derivation of the evolution equation of the correlation matrix}\label{appB}

In this section, we give a detailed derivation of Eq.~(\ref{eq:CX}). We reproduce Eq.~(\ref{eq:lindblad}), the quadratic Lindblad equation, as follows
\begin{equation}
\frac{d\hat{\rho}}{dt}=-i[\hat{H},\hat{\rho}]+\sum_{\mu}(2\hat{L}_{\mu}\hat{\rho}\hat{L}_{\mu}^{\dagger}-\{\hat{L}_{\mu}^{\dagger}\hat{L}_{\mu},\hat{\rho}\}),\label{eq:Lindblad}
\end{equation}
where
\begin{equation}
\hat{H}=\sum_{ij}\hat{c}_{i}^{\dagger}H_{ij}\hat{c}_{j},\hat{L}_{\mu}=\sum_{i}D_{\mu i}\hat{c}_{i}.\label{eq:HandL}
\end{equation}
For jump operators, we define the matrix $M$ as $M_{ij}=\sum_{\mu}D_{\mu i}^{*}D_{\mu j}$. Substituting the definition of the correlation matrix, $C_{ij}=\text{Tr}(\hat{c}_{i}^{\dagger}\hat{c}_{j}\hat{\rho})$, into Eq.~(\ref{eq:Lindblad}), and after rearranging the terms, we have
\begin{equation}
\frac{d{C}_{ij}}{dt}=i\text{Tr}([\hat{H},\hat{c}_{i}^{\dagger}\hat{c}_{j}]\hat{\rho})+\sum_{\mu}\text{Tr}[(2\hat{L}_{\mu}^{\dagger}[\hat{c}_{i}^{\dagger}\hat{c}_{j},\hat{L}_{\mu}]+[\hat{L}_{\mu}^{\dagger}\hat{L}_{\mu},\hat{c}_{i}^{\dagger}\hat{c}_{j}])\hat{\rho}].\label{eq:evoCij}
\end{equation}
Then we calculate each term on the right-hand side of Eq.~(\ref{eq:evoCij})
\begin{equation}
\begin{split}
[\hat{H},\hat{c}_{i}^{\dagger}\hat{c}_{j}]&=[\sum_{mn}\hat{c}_{m}^{\dagger}H_{mn}\hat{c}_{n},\hat{c}_{i}^{\dagger}\hat{c}_{j}]\\
&=\sum_{mn}H_{mn}[\hat{c}_{m}^{\dagger}\hat{c}_{n},\hat{c}_{i}^{\dagger}\hat{c}_{j}]\\
&=\sum_{mn}H_{mn}(\delta_{ni}\hat{c}_{m}^{\dagger}\hat{c}_{j}-\delta_{mj}\hat{c}_{i}^{\dagger}\hat{c}_{n})\\
&=\sum_{m}H_{mi}\hat{c}_{m}^{\dagger}\hat{c}_{j}-\sum_{n}H_{jn}\hat{c}_{i}^{\dagger}\hat{c}_{n},
\end{split}
\end{equation}
\begin{equation}
\begin{split}
\text{Tr}([\hat{H},\hat{c}_{i}^{\dagger}\hat{c}_{j}]\hat{\rho})&=\sum_{m}H_{mi}\text{Tr}(\hat{c}_{m}^{\dagger}\hat{c}_{j}\hat{\rho})-\sum_{n}H_{jn}\text{Tr}(\hat{c}_{i}^{\dagger}\hat{c}_{n}\hat{\rho})\\
&=\sum_{m}{H^T}_{mi}C_{mj}-\sum_{n}C_{in}{H^T}_{nj}\\
&={[H^T,C]}_{ij},\label{eq:term1}
\end{split}
\end{equation}
\begin{equation}
\begin{split}
\sum_{\mu}\hat{L}_{\mu}^{\dagger}[\hat{c}_{i}^{\dagger}\hat{c}_{j},\hat{L}_{\mu}]&=\sum_{\mu}\sum_{m}D_{\mu m}^{*}\hat{c}_{m}^{\dagger}[\hat{c}_{i}^{\dagger}\hat{c}_{j},\sum_{n}D_{\mu n}\hat{c}_{n}]\\
&=\sum_{\mu mn}D_{\mu m}^{*}D_{\mu n}\hat{c}_{m}^{\dagger}[\hat{c}_{i}^{\dagger}\hat{c}_{j},\hat{c}_{n}]\\
&=-\sum_{\mu mn}D_{\mu m}^{*}D_{\mu n}\delta_{in}\hat{c}_{m}^{\dagger}\hat{c}_{j}\\
&=-\sum_{\mu m}D_{\mu m}^{*}D_{\mu i}\hat{c}_{m}^{\dagger}\hat{c}_{j},
\end{split}
\end{equation}
\begin{equation}
\begin{split}
\text{Tr}(\sum_{\mu}\hat{L}_{\mu}^{\dagger}[\hat{c}_{i}^{\dagger}\hat{c}_{j},\hat{L}_{\mu}])&=-\sum_{\mu m}D_{\mu m}^{*}D_{\mu i}\text{Tr}(\hat{c}_{m}^{\dagger}\hat{c}_{j}\hat{\rho})\\
&=\sum_{m}M_{mi}C_{mj}\\
&=(M^T C)_{ij},\label{eq:term2}
\end{split}
\end{equation}
\begin{equation}
\begin{split}
\sum_{\mu}[\hat{L}_{\mu}^{\dagger}\hat{L}_{\mu},\hat{c}_{i}^{\dagger}\hat{c}_{j}]&=\sum_{\mu}[\sum_{m}D_{\mu m}^{*}\hat{c}_{m}^{\dagger}\sum_{n}D_{\mu n}\hat{c}_{n},\hat{c}_{i}^{\dagger}\hat{c}_{j}]\\
&=\sum_{\mu mn}D_{\mu m}^{*}D_{\mu n}[\hat{c}_{m}^{\dagger}\hat{c}_{n},\hat{c}_{i}^{\dagger}\hat{c}_{j}]\\
&=\sum_{\mu mn}D_{\mu m}^{*}D_{\mu n}(\delta_{ni}\hat{c}_{m}^{\dagger}\hat{c}_{j}-\delta_{mj}\hat{c}_{i}^{\dagger}\hat{c}_{n})\\
&=\sum_{\mu m}D_{\mu m}^{*}D_{\mu i}\hat{c}_{m}^{\dagger}\hat{c}_{j}-\sum_{\mu n}D_{\mu j}^{*}D_{\mu n}\hat{c}_{i}^{\dagger}\hat{c}_{n},
\end{split}
\end{equation}
\begin{equation}
\begin{split}
&\text{Tr}(\sum_{\mu}[\hat{L}_{\mu}^{\dagger}\hat{L}_{\mu},\hat{c}_{i}^{\dagger}\hat{c}_{j}]\hat{\rho})\\
&=\sum_{\mu m}D_{\mu m}^{*}D_{\mu i}\text{Tr}(\hat{c}_{m}^{\dagger}\hat{c}_{j}\hat{\rho})-\sum_{\mu n}D_{\mu j}^{*}D_{\mu n}\text{Tr}(\hat{c}_{i}^{\dagger}\hat{c}_{n}\hat{\rho})\\
&=\sum_{m}M_{mi}C_{mj}-\sum_{n}M_{jn}C_{in}\\
&=(M^{T} C)_{ij}-(C {M}^{T})_{ij}.\label{eq:term3}
\end{split}
\end{equation}
Substituting Eqs.~(\ref{eq:term1}), (\ref{eq:term2}), and (\ref{eq:term3}) into Eq.~(\ref{eq:evoCij}), we have
\begin{equation}
\frac{d{C}_{ij}}{dt}=i{[H^T ,C]}_{ij}-{\{ M^T ,C\}}_{ij}.\label{eq:EvoCij}
\end{equation}
Defining the damping matrix $X=i{H}^{T}-{M}^{T}$ and substituting $X$ into Eq.~(\ref{eq:EvoCij}), we get Eq.~(\ref{eq:CX}) of the main text
\begin{equation}
\frac{dC}{dt}=XC+C{X}^{\dagger}.\label{eq:EvoC}
\end{equation}

\renewcommand{\thefigure}{C\arabic{figure}}
\renewcommand{\thetable}{C\Roman{table}}
\setcounter{figure}{0}
\renewcommand{\theequation}{C\arabic{equation}}
\setcounter{equation}{0}

\section{Preserving the chiral symmetry}\label{appC}

We start by showing that the presence of the chiral symmetry is equivalent to $\bm{n}_{K}$ lying in the plane perpendicular to $\bm{n}_\Gamma$. From Eq.~(\ref{eq:chiral}) of the main text, the chiral symmetry is given by
\begin{equation}
\Gamma (\bm{n}_{K}\cdot \bm{\sigma})\Gamma^{-1}=-\bm{n}_{K}\cdot \bm{\sigma},\label{eq:Chiral}
\end{equation}
where the symmetry operator
\begin{equation}
\Gamma=\bm{n}_\Gamma\cdot \bm{\sigma},\label{eq:chiop}
\end{equation}
with the unit vector $\bm{n}_\Gamma$ denoting the chiral symmetry axis. Substituting Eq.~(\ref{eq:chiop}) into Eq.~(\ref{eq:Chiral}), we have
\begin{equation}
(\bm{n}_\Gamma \cdot \bm{\sigma})(\bm{n}_K\cdot \bm{\sigma})=-(\bm{n}_K\cdot \bm{\sigma})(\bm{n}_\Gamma \cdot \bm{\sigma}).\label{eq:chiralop}
\end{equation}
Consider the formula
\begin{equation}
(\bm{m}_1 \cdot \bm{\sigma})(\bm{m}_2 \cdot \bm{\sigma})=(\bm{m}_1 \cdot \bm{m}_2)\bm{\sigma}+i(\bm{m}_1 \times \bm{m}_2) \cdot \bm{\sigma},\label{eq:formula}
\end{equation}
where $\bm{m}_1$ and $\bm{m}_2$ are three-dimensional vectors. Substituting Eq.~(\ref{eq:formula}) into Eq.~(\ref{eq:chiralop}), we have
\begin{equation}
(\bm{n}_{\Gamma} \cdot \bm{n}_{K}) =0,
\end{equation}
which means all $\bm{n}_{K}$ (for all $k$) lie in the plane perpendicular to $\bm{n}_\Gamma$.

In the main text, we give a sufficient condition, in terms of $\tilde{X}_{k}$, for the preservation of the chiral symmetry of $C^T$ throughout the time evolution. Here we give a detailed derivation. Substituting Eqs.~(\ref{eq:ptXk}) and (\ref{eq:Ck}) into Eq.~(\ref{eq:kspaceCX}), we have
\begin{align}
\frac{d}{dt}\alpha_{C}&=2\left[\alpha_{X}\alpha_{C}+\text{Re}(\bm{n}_{C}\cdot \bm{n}_{X})\right],\label{eq:alphaCk}\\
\frac{d}{dt}\bm{n}_{C}&=2\left[\alpha_{X}\bm{n}_{C}+\alpha_{C}\text{Re}\bm{n}_{X}+\text{Im}(\bm{n}_{C}\times\bm{n}_{X})\right].\label{eq:nCk}
\end{align}
At $t=0$, we assume that all $\bm{n}_{C}$ lie in the same plane, so that the chiral symmetry is present initially.
We label the plane as $\gamma_{\Gamma}$. Then, for the chiral symmetry to hold at any given time,
the form of $\tilde{X}_{k}$ should be such that all $\bm{n}_{C}$ remain in the plane $\gamma_{\Gamma}$ throughout the time evolution.

Under the condition that $\text{Im}(\bm{n}_{X})$ is perpendicular to $\gamma_{\Gamma}$, from Eqs.~(\ref{eq:ptXk}) and (\ref{eq:ptnXk}), $\text{Re}(\bm{n}_{X})$ is located within the plane $\gamma_{\Gamma}$.
We can find that on the right-hand side of Eq.~(\ref{eq:nCk}), if all $\bm{n}_{C}(t)$ are in $\gamma_{\Gamma}$, then the vectors $\alpha_{X}(t)\bm{n}_{C}(t)$, $\alpha_{C}(t)\text{Re}\bm{n}_{X}(t)$ and $\text{Im}(\bm{n}_{C}(t)\times\bm{n}_{X}(t))$ are all in the plane $\gamma_{\Gamma}$, which
means that $\bm{n}_{C}(t)$ remains within the same plane at any given time $t$.

In the main text, we also point out that $\bm{n}_{C}$ and $\bm{n}_{K}$ are parallel but opposite in direction for all $k$, here we provide a detailed explanation. Using Eq.~(\ref{eq:Kk}), we have
\begin{align}
e^{K_k}=e^{\alpha_K}[\cosh(|\bm{n}_K|)+\sinh(|\bm{n}_K|)\frac{\bm{n}_K}{|\bm{n}_K|}\cdot \bm{\sigma}].\label{eq:eKk}
\end{align}
Substituting Eq.~(\ref{eq:eKk}) into Eq.~(\ref{eq:kspaceCK}), we have
\begin{align}
C^T_k=\frac{(e^{\alpha_K}\cosh(|\bm{n}_K|)+1)\sigma_0-e^{\alpha_K}\sinh(|\bm{n}_K|)\frac{\bm{n}_K}{|\bm{n}_K|}\cdot \bm{\sigma}}
{{(e^{\alpha_K}\cosh(|\bm{n}_K|)+1)}^2-{(e^{\alpha_K}\sinh(|\bm{n}_K|))}^2}.\label{eq:Cksigma}
\end{align}
Comparing Eq.~(\ref{eq:Cksigma}) and Eq.~(\ref{eq:Ck}), we have
\begin{align}
\bm{n}_{C}=-\frac{e^{\alpha_K}\sinh(|\bm{n}_K|)\frac{\bm{n}_K}{|\bm{n}_K|}\cdot \bm{\sigma}}
{{(e^{\alpha_K}\cosh(|\bm{n}_K|)+1)}^2-{(e^{\alpha_K}\sinh(|\bm{n}_K|))}^2}.\label{eq:nCsigma}
\end{align}
Based on Eq.~(\ref{eq:nCsigma}), and noting that $\alpha_{K}$ is a real number, we conclude that $\bm{n}_{C}$ and $\bm{n}_{K}$ are opposite in direction for all $k$.

\renewcommand{\thefigure}{D\arabic{figure}}
\renewcommand{\thetable}{D\Roman{table}}
\setcounter{figure}{0}
\renewcommand{\theequation}{D\arabic{equation}}
\setcounter{equation}{0}

\section{Evolution of $C^T_k$ in the form of left and right eigenstates}\label{appD}

In this section, we give a detailed derivation of Eq.~(\ref{eq:Ct}). Consider the PT symmetric $\tilde{X}_{k}$ in the form of left and right eigenvectors
\begin{equation}
\tilde{X}_{k}=\sum_{\xi=\pm} \varepsilon_{\xi k}|R_{\xi k} \rangle \langle L_{\xi k}|.\label{eq:diracX}
\end{equation}
Using the biorthonormal condition, the corresponding time-evolution operator can be written as
\begin{equation}
e^{\tilde{X}_{k} t}=\sum_{\xi=\pm} e^{\varepsilon_{\xi k} t}|R_{\xi k} \rangle \langle L_{\xi k}|.\label{eq:diraceX}
\end{equation}
To find the solution of Eq.~(\ref{eq:kspaceCX}), we first define
\begin{equation}
\tilde{C}^T_k(t)=e^{-\tilde{X}_{k} t} C^T_{k}(t) e^{-\tilde{X}_k^{\dagger} t}.\label{eq:tildeC}
\end{equation}
Taking the time derivative on both sides of Eq.~(\ref{eq:tildeC}), and noting that $\tilde{X}_{k}$ is time independent, we have
\begin{equation}
\frac{d}{dt}\tilde{C}^T_k(t)=-\tilde{X}_{k} \tilde{C}^T_{k}(t)-\tilde{C}^T_{k}(t) \tilde{X}_k^{\dagger}+e^{-\tilde{X}_{k} t} \frac{d}{dt}C^T_{k}(t) e^{-\tilde{X}_k^{\dagger} t}.\label{eq:dtildeC}
\end{equation}
Substituting Eq.~(\ref{eq:kspaceCX}) into Eq.~(\ref{eq:dtildeC}), we have
\begin{equation}
\frac{d}{dt}\tilde{C}^T_k(t)=0,\label{eq:solvetildeC}
\end{equation}
leading to
\begin{equation}
\tilde{C}^T_k(t)=\tilde{C}^T_k(0).\label{eq:solutildeC}
\end{equation}
Substituting Eq.~(\ref{eq:tildeC}) into Eq.~(\ref{eq:solutildeC}), we find the solution of Eq.~(\ref{eq:kspaceCX})
\begin{equation}
C^T_{k}(t)=e^{\tilde{X}_{k} t} C^T_{k}(0) e^{\tilde{X}_k^{\dagger} t}.\label{eq:solCk}
\end{equation}
Finally, substituting Eq.~(\ref{eq:diraceX}) into Eq.~(\ref{eq:solCk}), we arrive at Eq.~(\ref{eq:Ct}) of the main text.

\end{document}